\documentclass[11pt,reprint,nocopyrightspace]{sigplanconf-pldi16}

\usepackage{etex}
\usepackage[usenames,dvipsnames,svgnames,table]{xcolor}
\usepackage{listings}
\usepackage{pifont}
\usepackage{multirow}
\usepackage{amsmath}
\usepackage{amsfonts}
\usepackage{amssymb}
\usepackage{xspace}
\usepackage{graphicx}
\usepackage{lastpage}
\usepackage{paralist}
\usepackage{fancyhdr}
\usepackage{wrapfig}
\usepackage{balance}
\usepackage{tikz}
\usepackage{ctable} 
\usepackage{pgfplots}
\usepackage{amsthm}
\usepackage{dsfont}
\usepackage{colonequals}
\usepackage{enumitem}
\usepackage{wasysym}
\usepackage{mathtools}
\usepackage{listings}
\usepackage{algorithm}
\usepackage{algpseudocode}
\usepackage{pifont}
\usepackage{stmaryrd}
\usepackage{comment}
\usepackage[hyphens]{url}
\usepackage{etoolbox}
\usepackage{pbox}
\usepackage{relsize}
\usepackage{csquotes}
\usepackage{scalerel}
\usepackage{setspace}
\usepackage[T1]{fontenc}
\usepackage{manfnt}
\usepackage{tabularx}
\usepackage{hhline}
\usepackage{titlesec}
\usepackage{enumitem}

\DeclareFontFamily{U}{mathx}{\hyphenchar\font45}
\DeclareFontShape{U}{mathx}{m}{n}{
      <5> <6> <7> <8> <9> <10>
      <10.95> <12> <14.4> <17.28> <20.74> <24.88>
      mathx10
      }{}
\DeclareSymbolFont{mathx}{U}{mathx}{m}{n}
\DeclareFontSubstitution{U}{mathx}{m}{n}
\DeclareMathAccent{\widecheck}{0}{mathx}{"71}
\DeclareMathAccent{\wideparen}{0}{mathx}{"75}

\newtoggle{long}

\theoremstyle{definition}

\AtEndEnvironment{example}{\null\hfill\qedsymbol}%
\theoremstyle{plain}

\algrenewcommand\alglinenumber[1]{\smaller #1:}
\makeatletter
\lst@InstallKeywords k{attributes}{attributestyle}\slshape{attributestyle}{}ld
\makeatother

\usepackage[scaled=0.80]{DejaVuSansMono}
\lstdefinestyle{customc}{
  belowcaptionskip=1\baselineskip,
  breaklines=true,
  xleftmargin=\parindent,
  language=C,
  showstringspaces=false,
  basicstyle=\small\ttfamily,
  keywordstyle=\small\bfseries\ttfamily\color{NavyBlue},
  commentstyle=\itshape\color{black},
  identifierstyle=\ttfamily\color{black},
  stringstyle=\itshape\color{NavyBlue},
  keywords={ map, flatMap, reduce, return, define, gaussian, gauss, step, elif, then, skip, if, else, reduceByKey, sensitiveAttribute, fairnessTarget, def},
moreattributes={ let, where}, 
attributestyle = \itshape\ttfamily\color{attributecolor}
}

\setlist[itemize]{
    topsep=.5ex,
    itemsep=0ex,
    leftmargin=1em,
}

\setlist[description]{
    labelindent=.4cm,
    style=unboxed,
    leftmargin=.4cm,
    font=\itshape,
    topsep=.5ex,
    itemsep=1ex
}

\everypar{\looseness=-1}

\definecolor{lightgray}{gray}{0.9}
\definecolor{midgray}{gray}{0.65}
\definecolor{darkgray}{gray}{0.3}

\newcommand{\abr}[1]{\textsc{\MakeLowercase{#1}}}

\newcommand{\abrs}[1]{\abr{#1}{\footnotesize{s}}\xspace}

\renewcommand{\vec}[1]{\boldsymbol{#1}}

\renewcommand{\qedsymbol}{{\footnotesize{$\blacksquare$}}}

\newcommand{\rone}{(\emph{i})~}
\newcommand{\rtwo}{(\emph{ii})~}
\newcommand{\rthree}{(\emph{iii})~}

\definecolor{keywordcolor}{gray}{0.0}
\definecolor{attributecolor}{gray}{0.0}
\definecolor{wildcolor}{gray}{0.8}

\usepackage{relsize}

\newcommand{\true}{\emph{true}}

\newcommand\lin{\lstinline[mathescape,basicstyle=\ttfamily]}
\newcommand{\linn}[1] {\begin{center}\lin{1}\end{center}}

\newcommand{\dist}{D}

\newcommand{\pop}{\dist_{\emph{pop}}}
\newcommand{\pr}[1]{\text{Pr}[#1]}

\newcommand{\prog}{\mathcal{P}}

\newcommand{\fs}{FairSquare\xspace}

\pgfmathdeclarefunction{gauss}{2}{%
  \pgfmathparse{1/(#2*sqrt(2*pi))*exp(-((x-#1)^2)/(2*#2^2))}%
}

\let\originalleft\left
\let\originalright\right
\renewcommand{\left}{\mathopen{}\mathclose\bgroup\originalleft}
\renewcommand{\right}{\aftergroup\egroup\originalright}

\togglefalse{long}

\begin{document}

\title{Fairness as a Program Property}{}

\authorinfo{Aws Albarghouthi, Loris D'Antoni, Samuel Drews}{University of Wisconsin--Madison}{}
\authorinfo{Aditya Nori}{Microsoft Research}{}

\maketitle

\noindent
\textbf{Abstract}  We explore the following question:
  \emph{Is a decision-making program fair,
  for some useful definition of fairness?}
  First, we describe how several \emph{algorithmic fairness}
  questions can be phrased as \emph{program verification problems}.
	Second, we discuss an
  automated verification technique for
  proving  or disproving  fairness of decision-making programs
  with respect to a \emph{model of the population}.

\section{Introduction}\label{sec:intro}


Algorithms have  become powerful arbitrators
of a range of significant decisions with far-reaching
societal impact---hiring~\cite{articleTimesHiring,articleWired},
welfare allocation~\cite{articleSlateWelfare},
prison sentencing~\cite{articlePropublica},
policing~\cite{articleGuardianCrime,perry2013predictive},
amongst many others.
%
%
%
%
With the range and sensitivity of algorithmic decisions expanding by the day,
the question of whether an algorithm is \emph{fair} is a pressing
one.
Indeed, the notion of \emph{algorithmic fairness} has captured
the attention of a broad spectrum of experts:
machine learning and theory researchers~\cite{dwork12,zemel13,feldman15,calders10}; privacy researchers and investigative journalists~\cite{datta2015automated,articlePropublica,articWsjstaples,sweeney2013discrimination};
law scholars and social scientists~\cite{tutt2016fda,ajunwa2016hiring,barocas2014big}; governmental agencies and \abrs{NGO}~\cite{articleWhitehouse14}.

Ultimately, algorithmic fairness is a question
about programs and their properties: \emph{Is a given
program $\prog$ fair, under some definition of fairness?
Or, how fair is $\prog$?}
In this paper, we describe a line of work
that approaches the question of algorithmic
fairness from a program-analytic perspective,
in which our goal is to \emph{analyze} a given
decision-making program and \emph{construct
a proof} of its fairness or unfairness---%
just as a traditional static program verifier would prove
correctness of a program with respect to, for example,
lack of divisions by zero, integer overflows,
null-pointer derefrences, etc.

We start by analyzing what are the challenges and research questions
in checking algorithmic fairness for decision making programs (Section~\ref{sec:proving}).
We then present a simple case study and show how techniques for verifying probabilistic
programs can be used to automatically prove or disprove global fairness for
a class of programs that subsume a range of machine learning classifiers (Section~\ref{sec:casestudy}).
Finally, we
lay a list of
many challenging and interesting questions that the algorithms and programming languages communities
need to answer to achieve the ultimate goal of building a fully automated system for verifying and guaranteeing algorithmic fairness in real-world applications (Section~\ref{sec:future}).

\begin{figure}[t]
 \centering
\includegraphics[scale=1.1]{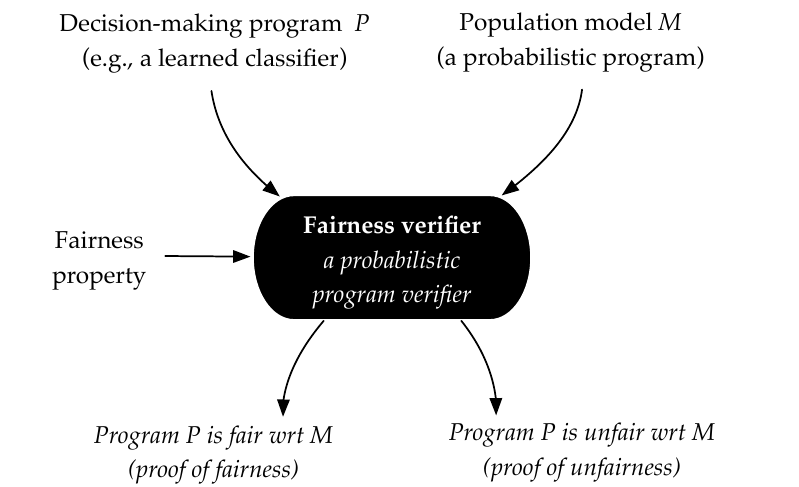}
\caption{Overview}
\label{fig:overview}
\end{figure}

\section{Proving Programs Fair}
\label{sec:proving}

In this section, we describe the  components of the fairness verification problem.
Intuitively, our goal is to prove whether a certain program is fair with respect to the set of possible inputs
over which it operates.
Tackling the fairness-verification problem
requires answering a number of challenging questions:
\begin{itemize}
\item What class of decision-making programs should our program model capture?
\item How can we define the set of possible inputs to the program and
capture complex
probability distributions that are useful and amenable to verification?
\item How can we describe what it means for the program to be fair?
\item How can we fully automate the  verification process?
\end{itemize}

Figure~\ref{fig:overview}
provides a high-level picture of our proposed framework.
As shown, the \emph{fairness verifier} takes a (white-box) decision-making
program $\prog$ and a population model $M$.
The verifier then proceeds to prove or disprove that $\prog$ is fair for the given population defined by the model $M$.
Here, the model $M$ defines a joint probability distribution on the inputs of $\prog$.
 Existing definitions of fairness define programs as fair or unfair with respect to a given concrete dataset. While using a concrete dataset simplifies the verification problem, it also raises questions of whether the dataset is representative for the population for which we are trying to prove fairness. Our technique moves away from concrete datasets and replaces them with a probabilistic population model.
 We envision a future in which fairness verification is regulated.\footnote{The European Union (EU), for instance, has already begun regulating algorithmic decision-making~\cite{Goodman16}.}
 For instance, a governmental agency can publish a probabilistic
 population model (e.g., generated from census data).
 Any organization employing a decision-making algorithm
 with potentially significant consequences (e.g., hiring) must quantify fairness of their algorithmic
 process against the current picture of the population, as specified by the population model.

\paragraph{Decision-making programs}
In the context of algorithmic fairness, a program $\prog$ takes as input a vector
of arguments $\vec{v}$ representing a set of input attributes (features),
where one (or more) of the arguments $v_s$ in the vector $\vec{v}$ is \emph{sensitive}---e.g., gender or race.
Evaluating $\prog(\vec{v})$ may return a Boolean value indicating---e.g., hire or not hire---if the program is a binary
or a numerical value---e.g., a mortgage rate.
The set of combinators, operations, and types used by the program
can vastly affect the complexity of the verification procedures.
For example, loops are the hardest type of programming construct to reason about, but
most machine learning classifiers do not contain loops.
Similarly, since classifiers typically operate over real values, we
can limit the set of possible types allowed in our programs to only being reals or
other types that can be \emph{desugared} into reals.
All these decisions are crucial in the design of a verification
procedure.

\paragraph{Population model}
To be able to reason about the outcome of the program we need
 to specify what kind of input the program will operate on.
For example, although a program that allocates mortgages might be ``fair'' with respect a certain set of  applicants, it may become
unfair when considering a different pool of people.
In program verification, the ``kind of inputs'' over which the program operates is called the precondition
and is typically stated as a formal logical property with the program inputs as free variables.
An example of program precondition is
$$v_{\emph{gender}}=f \rightarrow  v_{\emph{job}}\neq \emph{priest}$$
which indicates that none of the program inputs is both a woman and a priest. Of course, there are many possible choices for what language we can use to describe the program's precondition. In particular, if we want to capture a certain probability distribution over the input of the program, our language will be a logic that can describe probabilities and random variables. 
For example, we might want to be able to specify that half of the inputs are female, $\pr{v_{gender}=f}=0.5$,
or that the age of the processed inputs has a particular distribution, $v_{age} \sim gauss(18,5)$. 
Again, the choice of the language allowed in the preconditions is crucial in the design of a verification procedure.
From now on, we refer to the program precondition, $\pop$, as the \emph{population model}.

\paragraph{Fairness properties}
The next step is to define a property stating that the program's outcome is fair
with respect to the program's precondition.
In program verification, this is called the postcondition of the program.
As observed in the fairness literature, there are many ways to define
when and why a program is fair or unfair.

For example, if we want to prove \emph{group fairness}---i.e., that the algorithm is \emph{just as likely}
to hire a minority applicant ($m$) as it is for other, non-minority applicants---our postcondition will be an expression of the form
$$\frac{\pr{\prog(\vec{v}) = \true  \mid v_s=m}}
    {\pr{\prog(\vec{v}) = \true \mid  v_s\neq m}}
>
1 - \epsilon$$
where $\true$ is the desired return value of the program, e.g., indicating hiring.
On the other hand, if we want to prove \emph{individual fairness}---i.e., \emph{similar} inputs should have
similar outcomes---%
our postcondition will be an expression of the form
$$\pr{\prog(\vec{v})\neq \prog(\vec{v'}) \mid \vec{v}\sim \vec{v'}}<\epsilon$$
Notice that  the last postcondition
relates the outcomes of the program on different input values.
As the two types of properties we described are radically different, they will also require
different verification mechanisms.

\paragraph{Proofs of (un)fairness}
The task of proving whether a program is fair boils down to statically checking
whether, on inputs satisfying the precondition, the outcome of the program satisfies
the post-condition.
For simple definitions, such as group fairness,
the verification problem reduces to computing the probability
of a number of events with respect to the program
and the population model.
For more complex definitions, such as individual fairness,
proving fairness requires more complex reasoning involving multiple
runs of the programs (i.e., a \emph{hyperproperty}~\cite{clarkson2010hyperproperties}), a notoriously hard problem.
In the case of a negative result, the verifier should provide the users with a proof of  unfairness.
Depending on the fairness definition, producing a human-readable proof might be challenging
as the argument might involve multiple and potentially infinite inputs.
For example, in the case of group fairness it might be challenging to explain
why the program outputs true on 40\% of the minority inputs and on 70\% of the majority inputs.

\section{Case Study}
\label{sec:casestudy}

We now describe a simplified case study
demonstrating how our fairness verification
methodology can be used to prove or disprove
fairness of a given decision-making program.

\paragraph{A program and a population model}
Consider the following program \texttt{dec},
which is a decision-making
program that takes a job applicant's
college ranking and years of experience and
decides whether they get hired or not (the \emph{fairness target}).
The program implements a decision tree,
perhaps one generated by a machine-learning algorithm.
A person is hired if they attended a \emph{top-5}  college (\texttt{colRank <= 5}) or  have lots of experience compared to their college's ranking (\texttt{expRank > -5}).
Observe that \texttt{dec} \emph{does not access  ethnicity}.

\begin{lstlisting}
define dec(colRank, yExp)
  expRank $\gets$ yExp - colRank
  if (colRank <= 5)
    hire $\gets$ true
  elif (expRank > -5)
    hire $\gets$ true
  else
    hire $\gets$ false
  return hire
\end{lstlisting}

Now, consider the program \texttt{popModel}, which is a probabilistic
program describing a simple model of the population.
Here, a member of the population has three
attributes, all of which are real-valued:
\rone \texttt{ethnicity};
\rtwo \texttt{colRank}, the ranking of the college
the person attended (lower is better);
and \rthree \texttt{yExp}, the years of work experience
a person has.
We consider a person is a member of a protected group
if \texttt{ethnicity > 10}; we call this the
\emph{sensitive condition}.
The population model can be viewed as a \emph{generative model}
of records of individuals---the more likely a combination
is to occur in the population, the more likely it will be generated.
For instance, the years of experience an individual has (line 4)
follows a Gaussian
distribution with mean $10$ and standard deviation $5$.

\begin{lstlisting}
define popModel()
  ethnicity ~ gauss(0,10)
  colRank ~ gauss(25,10)
  yExp ~ gauss(10,5)
  if (ethnicity > 10)
    colRank $\gets$ colRank + 5
  return colRank, yExp
\end{lstlisting}

\paragraph{A note on the program model}
Note that our program model, while admitting arbitrary programs,
is rich enough to capture programs (classifiers) generated
by standard machine learning algorithms.
For example, linear support vector machines, decision
trees, and neural networks, can be represented in our language
simply using assignments with arithmetic expressions and conditionals.
Similarly, the population model is a probabilistic program, where assignments can be made
by drawing values from predefined distributions. Like other
probabilistic programming languages, our programming model
is rich enough to subsume graphical models like Bayesian networks~\cite{gordon2014probabilistic}.

\paragraph{Group fairness}
Suppose that our goal is to prove
group fairness, following the definition
of Feldman et al.~\cite{feldman15}:
\[
  \frac{\pr{\texttt{hire}  \mid \texttt{min}}}
   {\pr{\texttt{hire} \mid \neg \texttt{min}}}
>
1 - \epsilon
\]
where \texttt{min} is shorthand for the sensitive condition
\texttt{ethnicity > 10}.

\paragraph{Probabilistic inference as volume computation}
To prove (un)fairness of the decision-making
model with respect to the population,
we need to compute the probabilities appearing in
the group fairness ratio.
For illustration, suppose we are computing the
probability $\pr{\texttt{hire} \land \neg \texttt{min}}$.
We need to reason about the \emph{composition} of the two
programs, $\texttt{dec}\circ\texttt{popModel}$.
That is, we want to compute the probability that
\rone \texttt{popModel} generates a non-minority applicant,
and \rtwo \texttt{dec} hires that applicant.
To do so, we observe that every possible execution of the composition
$\texttt{dec}\circ\texttt{popModel}$
is uniquely characterized by the set of the three probabilistic choices
made by \texttt{popModel}.
In other words, every execution is characterized by a vector
$\vec{v} \in \mathds{R}^3$.

Thus, our goal is to compute the probability that
we draw a vector $\vec{v}$ that results in a minority applicant
being hired.
Probabilistic programming languages,
e.g., Church~\cite{goodman2012church}, R2~\cite{nori2014r2}, and Stan~\cite{carpenter2015stan}, employ \emph{approximate inference}
techniques, like \abr{MCMC}, which converge in the limit
but offer no guarantees on how far we are from
the exact result.
In our work, we consider \emph{exact inference}, which has primarily
received attention in the Bayesian network setting, and boils
down to solving a \#SAT instance~\cite{chavira2008probabilistic}.
In our setting, however, we are dealing with real-valued variables.

Using standard techniques from program analysis and verification,
we can characterize the set of all such vectors
as a formula $\varphi$, which is comprised of Boolean combinations
(conjunctions/disjunctions) of linear inequalities---since our
program only has linear expressions.
Geometrically, the formula $\varphi$ is a set of convex polyhedra in $\mathds{R}^n$.
Therefore, the probability   $\pr{\texttt{hire} \land \neg \texttt{min}}$
is the same as the probability of drawing a vector $\vec{v}$ that
lies inside of $\varphi$.
In other words, we are interested in the \emph{volume} of $\varphi$,
weighted by the probabilistic choices.
Formally:
$$ \pr{\texttt{hire} \land \neg \texttt{min}} = \int_{\varphi} p_ep_yp_c ~dedydc$$
where, e.g., $p_e$ is the \emph{probability density function}
of the distribution \texttt{gauss(0,10)}---the distribution
from which the value of \texttt{ethnicity} is drawn  in line 2 of \texttt{popModel}.

\begin{figure}[t]
 \centering
\includegraphics[scale=1.0]{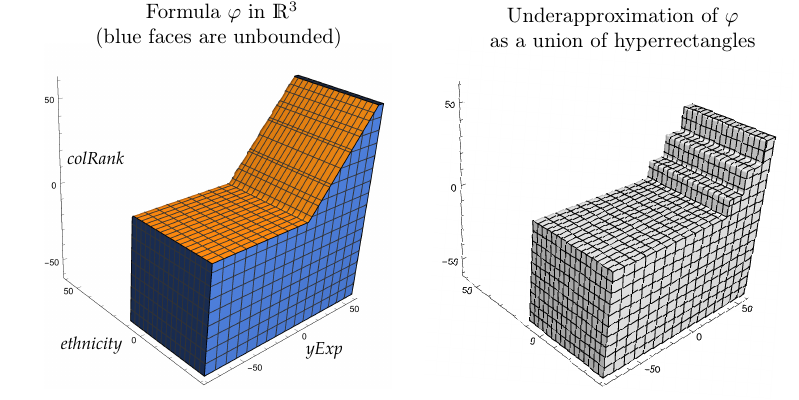}
\caption{Underapproximation of $\varphi$ as hyperrectangles}
\label{fig:ex}
\end{figure}

The volume computation problem is a well-studied and hard problem~\cite{khachiyan1993complexity,dyer1988complexity}.
Indeed, even for a convex polytope, computing its volume is \#P-hard.
Leveraging the great developments in \emph{satisfiabiltiy modulo theories}
(\abr{SMT}) solvers~\cite{barrett09}, we developed a procedure that reduces the volume
compuation problem to a series of calls to the \abr{SMT} solver, viewed
completely as an oracle.
Specifically, our procedure uses the \abr{SMT} solver to
\emph{sample} subregions of $\varphi$ that are \emph{hyperrectangular}.
Intuitively, for hyperrectangular regions in $\mathds{R}^n$, evaluating
the above integral is a matter of evaluating the \abrs{CDF} of the various
distributions.
Thus, by systematically sampling more and more non-overlapping hyperrectangles
in $\varphi$, we maintain a \emph{lower bound} on the probability of interest.
Figure~\ref{fig:ex} pictorially illustrates $\varphi$ and an under-approximation
with 4 hyperrectangles.
Similarly, to compute an \emph{upper bound} on the probability,
we can simply invoke our procedure on $\neg \varphi$.

\paragraph{Fairness certificates}
The fairness verification tool terminates
when it has computed lower/upper bounds that prove
or disprove the desired fairness criteria.
The hyperrectangles sampled in the process
of computing volumes can serve as proof certificates.
That is, an external entity can take the hyperrectangles,
compute their volumes, and ensure that they indeed lie
in the expected regions in $\mathds{R}^n$.

\section{Experience and future Outlook}
\label{sec:future}

\paragraph{Experience}
We have built a fairness-verification tool,
called \fs, that takes a decision-making
program, a population model, and verifies fairness
of the program with respect to the model.
So far, we have focused on group fairness.
The tool uses the popular Z3 \abr{SMT} solver~\cite{de2008z3}
for manipulating first-order formulas over arithmetic theories.

We have used \fs  to prove or disprove fairness of a
suite of population models and
programs representing
machine-learning classifiers that were automatically
generated from real-world datasets used in other
work on algorithmic fairness~\cite{feldman15,zemel13,datta2016algorithmic}.
Specifically, we have considered linear \abrs{SVM},
simple neural networks with rectified linear units,
and decision trees.

\paragraph{Future outlook}
Looking forward, we see a wide range
of avenues for improvement and exploration.
For instance, we are currently working on the problem
of \emph{making an unfair program fair}.
That is, given a program $\prog$ that is considered
unfair, what is the smallest \emph{tweak} that would make
it fair. Our goal is to \emph{repair} the program, making
it fair, while ensuring that it is semantically close
to the original program.

\vspace{.1in}

{
\bibliographystyle{plain}
\bibliography{biblio}
}

\end{document}